\documentclass[a4paper,11pt]{article}
\usepackage{graphicx}
\usepackage{amsmath}
\usepackage{amssymb}
\usepackage{latexsym}
\usepackage[colorlinks]{hyperref}
\usepackage{color}
\usepackage{cite}
\usepackage{float}
\usepackage[textfont={footnotesize,sf},labelfont={color=blue,bf,sf},labelsep=endash]{caption}
\usepackage[position=top,labelfont={color=blue,bf,sf}]{subfig}
\usepackage{bm}
\newcommand{\bra}{\begin{array}}
\newcommand{\era}{\end{array}}
\newcommand{\beq}{\begin{equation}}
\newcommand{\eeq}{\end{equation}}
\newcommand{\beqar}{\begin{eqnarray}}
\newcommand{\eeqar}{\end{eqnarray}}

\def\BC{\bb C}
\def\_\BC{\bbi C}

\def\( {\left(}
   \def\) {\right)}
\def\[ {\left[}
\def\] {\right]}
\def\no2 {{\textstyle{n\over 2}}}

\def\dag {{\dagger}}

\newcommand{\si}{\sigma}

\newcommand{\ov}{\over}

\newcommand{\lb}{\label}

\begin{document}
\begin{titlepage}
\setcounter{page}{1}
\renewcommand{\thefootnote}{\fnsymbol{footnote}}


\vspace{5mm}
\begin{center}

{\Large \bf {The correlation function of a two-dimensional electron gas with anharmonic potential and Rashba coupling}}

\vspace{5mm}

{\bf Rachid Hou\c{c}a$^{1,2}$\footnote{houca.rachid@gmail.com}
,
El Bouâzzaoui Choubabi$^{2}$\footnote{choubabi@gmail.com}
,
Abdellatif Kamal$^{2,3}$\footnote{abdellatif.kamal@ensam-casa.ma}
,
Abdelhadi Belouad$^{2}$\footnote{belabdelhadi@gmail.com}
and
Mohammed El Bouziani$^{2}$\footnote{elbouziani@yhoo.com}
}

\vspace{5mm}

{$^{1}$\em Team of Theoretical Physics and high energy, Department of Physics, Faculty of Sciences, Ibn
Zohr University}, Agadir, Morocco,\\
{\em PO Box 8106, Agadir, Morocco}

{$^{2}$\em L.P.M.C. Laboratory, Theoretical Physics Group,
Faculty of Sciences, Choua\"ib Doukkali University},\\
{\em PO Box 20, 24000 El Jadida, Morocco}

{$^{3}$\em Department of Mechanical Engineering, National Higher School of Arts and Crafts, Hassan II
University, Casablanca, Morocco}

\vspace{3cm}

\begin{abstract}
In this paper, we study a two-dimensional gas of electrons with Rashba spin-orbit coupling with anharmonic potential. The splitting of the Hamiltonian in two parts can be recovering the Jaynes-Cummings model, which describes a system with two states. The dynamics of the rising Pauli operator and the creation operators are studied. Finally, we examined the behavior of the correlation function of emission and absorption photons for the strong and weak coupling.
\end{abstract}
\end{center}
\vspace{5cm}


\end{titlepage}


\section{Introduction}
Spin electronics, or spintronics, is concerned with the degree of freedom offered by the electron's spin to create devices for storing, processing logic, or transporting information \cite{Zutic}. Since the $1990$s, two axes have developed in parallel: the spintronics of metals and semiconductors. Metal spintronics took off with the discovery of giant magnetoresistance in $1988$ \cite{Baibich,Binasch}. This discovery, which earned A. Fert and P. Grünberg the Nobel Prize in 2007, demonstrated that electronic transport could be controlled by the magnetization of ferromagnetic materials, itself controllable by a magnetic field. A direct application is the spin valve \cite{Dieny}.

Semiconductor spintronics \cite{Awschalom} is in line with the invention of the transistor in 1948. The transistor exploits the possibility, specific to semiconductors, of controlling the electronic properties by doping and applying electric fields. Semiconductors also have certain advantages for spintronics: the spin is easily accessible optically. Generally, it has a greater coherence than in metals. Thus, it was quickly considered to reproduce in semiconductors the spintronic effects initially demonstrated in metals. Interest has turned to hybrid ferromagnetic/semiconductor structures \cite{Ohno0,Crooker}, and to multifunctional materials, both semiconductor and magnetic diluted magnetic semiconductors like $GaMnAs$ \cite{Ohno1,Ohno2} or $CdMnTe$ \cite{Furdyna}. A third strategy has also emerged, which consists of exploiting the spin-orbit interaction of semiconductors.

The spin-orbit interaction is a relativistic effect. Whereby an electron was moving in an electric field sees an effective magnetic field $\vec{B}_{SO}$, which acts on its spin. This very general mechanism makes it possible to generate and manipulate the spin of electrons by simply influencing their movement by electric fields. It, therefore, opens up the possibility of avoiding the use of ferromagnetic materials or external magnetic fields and of considering all-electric semiconductor devices for spintronics. Thus, it has recently been shown that one could generate an accumulation of spin. By the simple passage of a charge current: this is the spin Hall effect \cite{Perel,Myers}. It has also been shown that continuous or pulsed electric fields can carry out coherent rotations of the spin of traveling electrons \cite{Kato,Kuhlen}.

Moreover, the Jaynes–Cummings (J-C) model makes it possible to represent the light-matter interaction insofar as it models a system made up, within an optical cavity, by an atom at two levels interacting with a quantized mode of the electromagnetic field. The model offers the advantage of lending itself to calculations of exact solutions. In addition, the simplicity of the representation of the interaction does not affect its ability to deal with a large number of observed phenomena. The appropriate framework for the formulation and practical testing of the J-C concept is quantum  electrodynamics cavity (QEC), as the fundamental experiments have been carried out through the use of microwaves and optical cavities \cite{Mabuchi,Rempe,Brune}. The use of semiconductor quantum dots in photonic crystal, micro-pillar, or micro-disk resonators \cite{Khitrova}, which may be used to investigate the domain of electrodynamic quantum cavities, is common in solid state systems as well.

Motivated by our research \cite{houca}, we broadened our previous approach to address a variety of problems. To do this, we examine a two-coupled harmonic oscillator operating in a uniform magnetic field $B$ and with Rashba spin-orbit coupling in order to establish a link with quantum optics and to investigate its various quantum dynamics. We diagonalize our system using the creation and annihilation operators to arrive at the solutions to the energy spectrum. We demonstrate that our system simplifies the Landau issue in the first instance by examining a strong and a weak $B$. We constructed two replicas of the Jaynes–Cummings model oscillating at different frequencies using the Heisenberg image.
The following is the structure of the current paper. In Section \textcolor[rgb]{0.00,0.07,1.00}{2}, we formalize our issue by specifying the Hamiltonian and selecting a suitable gauge for consideration. In section \textcolor[rgb]{0.00,0.07,1.00}{3}, as a result, we add a sequence of annihilation and creation operators to diagonalize our system. This allows us to calculate the precise eigenvalues and eigenstates of our Hamiltonian, which is useful for many applications.  In section \textcolor[rgb]{0.00,0.07,1.00}{4}, We build a connection with the Jaynes–Cummings model and make a variety of findings. In section \textcolor[rgb]{0.00,0.07,1.00}{5}, We will investigate the correlation function for photon absorption or emission, and we will analyze it for two limiting instances in order to demonstrate the significance of our findings. Finally, we give our conclusions in section \textcolor[rgb]{0.00,0.07,1.00}{6}.

\section{System energy}
We begin by defining our issue in order to arrive at the Hamiltonian that best describes the system under discussion. Indeed, when we examine a two-dimensional gas of electrons with Rashba spin-orbit coupling \cite{houca} and an anharmonic potential \cite{han}, the system's Hamiltonian reads as
\beq\lb{HH}
H={p_x^2\ov2m_1}+{p_y^2\ov2m_2}+{k_1\ov2}x^2+{k_2\ov2}y^2+{k_3\ov2}xy+\alpha\left(\sigma_yp_x-\sigma_xp_y\right),
\eeq
where $\alpha$  is the Rashba coupling parameter and $\vec{\si}=\left(\si_x,\si_y,\si_z\right)$ are the  Pauli matrices.
It is easy to incorporate additional phase space variables by rescaling those found in (\ref{HH}). Indeed, the positions may be defined as:
\beq\lb{pp}
P_x=\left({m_2\ov m_1}\right)^{1\ov4}p_x, \quad  P_y=\left({m_1\ov m_2}\right)^{1\ov4}p_y
\eeq
By substituting them all, we demonstrate that (\ref{HH}) becomes
\beq\lb{HHh}
H={P_x^2+P_y^2\ov2\mu}+{1\ov2}\left(k_1x^2+k_2y^2+k_3xy\right)+\alpha\left(\sigma_yp_x-\sigma_xp_y\right),
\eeq
In this case, $\mu=\sqrt{m_1m_2}$. Due to the fact that the Hamiltonian contains an interaction term, it is difficult to conduct a simple examination of the system's fundamental properties. Nonetheless, this issue may be simplified via a translation to new phase space variables
\beq\lb{tra}
\left(
  \begin{array}{c}
    P_x \\
    P_y \\
  \end{array}
\right)=M\left(
           \begin{array}{c}
             p_1 \\
             p_2 \\
           \end{array}
         \right); \qquad \left(
  \begin{array}{c}
    x \\
    y \\
  \end{array}
\right)=M\left(
           \begin{array}{c}
             q_1 \\
             q_2 \\
           \end{array}
         \right)
\eeq
Where $M$ is a unitary rotation matrix with a mixing angle, $\theta$ can be defined by:
\beq\lb{M}
M=\left(
  \begin{array}{cc}
    \cos{\theta\ov2} & -\sin{\theta\ov2} \\
    \sin{\theta\ov2} & \cos{\theta\ov2} \\
  \end{array}
\right)
\eeq
By including the mapping (\ref{M}) into (\ref{HHh}), it becomes clear that $\theta$ must fulfill the requirement
\beq
\tan\theta={k_3\ov k_2-k_1}
\eeq
To get the following Hamiltonian
\beq\lb{HHH}
H={1\ov2\mu}\left(p_1^2+p_2^2\right)+{k\ov2}\left(e^{2\lambda}q_1^2+e^{-2\lambda}q_2^2\right)+
\alpha\left[\left(\cos{\theta\ov2}\sigma_y-\sin{\theta\ov2}\sigma_x\right)p_1-\left(\sin{\theta\ov2}\sigma_y+\cos{\theta\ov2}\sigma_x\right)p_2\right]
\eeq
where
\beq
k=\sqrt{k_1k_2-{k_3^2/4}},\qquad e^\lambda={k_1+k_2+\sqrt{\left(k_1-k_2\right)^2+k_3^2}\ov 2k}.
\eeq
In addition, the condition $4k_1k_2 > k_3^2$ must be met. With the separation of Cartesian variables technique, it is simple to see the algebraic structure of the Hamiltonian described above. This method indicates that (\ref{HHH}) should be divided into two parts
\beq\lb{H'}
H=H_F+H_R
\eeq
As a result, the free part has the structure
\beq\lb{free}
H_F={1\ov2\mu}\left(p_1^2+p_2^2\right)+{k\ov2}\left(e^{2\lambda}q_1^2+e^{-2\lambda}q_2^2\right)
\eeq
as well as the Rashba coupling with the transformation above, which is given by:
\beq\lb{Ra}
H_R=\alpha\left[\left(\cos{\theta\ov2}\sigma_y-\sin{\theta\ov2}\sigma_x\right)p_1-\left(\sin{\theta\ov2}\sigma_y+\cos{\theta\ov2}\sigma_x\right)p_2\right].
\eeq
Specifically, we point out that (\ref{H'}) is divided into two separate harmonic oscillator Hamiltonians, with the Rashba spin–orbit coupling serving as a supplement. This handy version of the Hamiltonian will assist us in easily obtaining its diagonalization in a straightforward manner.
\section{Weyl–Heisenberg symmetry solution}
To acquire solutions to the energy spectrum of the Hamiltonian (\ref{H'}), we present conventional methods that have been developed throughout time. It is preferable to use annihilation operators on the oscillator instead of directly utilizing them
\beq
a_1={q_1\ov \ell}e^\lambda+{i \ell\ov 2\hbar}p_1,\quad a_2={q_2\ov \ell}e^{-\lambda}+{i \ell\ov 2\hbar}p_2
\eeq
where
\beq
[a_i,a_j^\dag]=e^\lambda\delta_{ij}.
\eeq
Currently, we deal with two additional ones, which are the linear combination of $a_1$ and $a_2$, such that
\begin{eqnarray}
  a_d &=& {1\ov\sqrt{2}}\left(a_1-ia_2\right) \\ \nonumber
  a_g &=& {1\ov\sqrt{2}}\left(a_1+ia_2\right)
\end{eqnarray}
where $\ell = \sqrt{{2\hbar\omega\ov k}}$ denotes the magnetic length and $\omega=\sqrt{{k\ov \mu}}$ represents the frequency of the new system. Note that bosonic operators $a_d$ and $a_g$ fulfill the relationship commutations
\beq
[a_g,a_g^\dag]=[a_d,a_d^\dag]=e^\lambda
\eeq
as well as the disappearance of all other relationships. It is possible to derive useful identities for momentum from the operators listed above
\begin{eqnarray}
  p_1 &=& {\hbar\ov i\sqrt{2}\ell}\left(a_d-a_d^\dag+a_g-a_g^\dag\right) \\ \nonumber
  p_2 &=& {\hbar\ov \sqrt{2}\ell}\left(a_d+a_d^\dag-a_g-a_g^\dag\right)
\end{eqnarray}
Also, for the positions
\begin{eqnarray}
  q_1 &=& {\ell\ov 2\sqrt{2}}\left(a_d+a_d^\dag+a_g+a_g^\dag\right) \\ \nonumber
  q_2 &=& {\ell \ov 2\sqrt{2}i}\left(-a_d+a_d^\dag+a_g-a_g^\dag\right)
\end{eqnarray}
These algebraic structures will be fundamental in resolving various problems and, more specifically, in diagonalizing the many Hamiltonians that are derived. We begin by simplifying the formulation of the Rashba Hamiltonian (\ref{Ra}), which is the first step
\beq
H_R=\alpha\left(
      \begin{array}{cc}
        0 & -e^{-i{\theta\ov2}}\left(p_2+ip_1\right) \\
        e^{i{\theta\ov2}}\left(-p_2+ip_1\right) & 0 \\
      \end{array}
    \right)
\eeq
where the phases appearing in the formula of $H_R$ are defined by
\beq
e^{\pm i{\theta\ov2}}=\cos{\theta\ov2}\pm i\sin{\theta\ov2}.
\eeq
 In terms of the annihilation and creation operators discussed above, the Hamiltonian $H_ R$ may be expressed as follows:
\beq
H_R=\sqrt{2}\alpha{\hbar\ov \ell}\left(
      \begin{array}{cc}
        0 & -e^{-i{\theta\ov2}}\left(a_d-a_g^\dag\right) \\
        e^{i{\theta\ov2}}\left(-a_d^\dag+a_g\right) & 0 \\
      \end{array}
    \right)
\eeq
The Hamiltonian in the above paragraph may be divided into two commuted parts. As a result, we have
\beq
H_R = H_R^d + H_R^g
\eeq
where the first and second parts are given by:
\begin{eqnarray}
  H_R^g &=& \sqrt{2}\alpha{\hbar \ov \ell}\left(
        \begin{array}{cc}
          0 & e^{-i{\theta\ov2}}a_g^\dag \\
          e^{i{\theta\ov2}}a_g & 0 \\
        \end{array}
      \right) \\
  H_R^d &=& -\sqrt{2}\alpha{\hbar\ov \ell}\left(
        \begin{array}{cc}
          0 & e^{-i{\theta\ov2}}a_d \\
          e^{i{\theta\ov2}}a_d^\dag & 0 \\
        \end{array}
      \right)
\end{eqnarray}
We split the total Hamiltonian into two commuted parts to find energy spectrum solutions for this issue
\beq\lb{dg}
H=H_d+H_g
\eeq
where the  Hamiltonians $H_g$ and $H_d$ are described by the formula
\begin{eqnarray}
  H_g &=& \left(
        \begin{array}{cc}
          \hbar\omega\left(a_g^\dag a_g+{1\ov2}\cosh \lambda\right) & \sqrt{2}\alpha{\hbar\ov l}e^{-i{\theta\ov2}}a_g^\dag \\
          \sqrt{2}\alpha{\hbar\ov l}e^{i{\theta\ov2}}a_g & \hbar\omega\left(a_g^\dag a_g+{1\ov2}\cosh \lambda\right) \\
        \end{array}
      \right) \\
  H_d &=& \left(
        \begin{array}{cc}
          \hbar\omega\left(a_d^\dag a_d+{1\ov2}\cosh \lambda\right) & -\sqrt{2}\alpha{\hbar\ov l}e^{-i{\theta\ov2}}a_d \\
          -\sqrt{2}\alpha{\hbar\ov l}e^{i{\theta\ov2}}a_d^\dag & \hbar\omega\left(a_d^\dag a_d+{1\ov2}\cosh \lambda\right) \\
        \end{array}
      \right)\lb{dg}
\end{eqnarray}
Where the terms in the diagonal represent the components of free Hamiltonian part $H_F$. To find the solutions of the energy spectrum for the Hamiltonian (\ref{dg}), we resolve the eigenvalues equations
\beq
H_d|\psi_{d}\rangle=E_d|\psi_d\rangle,\quad H_g|\psi_{g}\rangle=E_g|\psi_g\rangle
\eeq
where $|\psi_{d}\rangle$ and $|\psi_{g}\rangle$ are the $n_d$ and $n_g-$th Landau levels, respectively. The overall value of the system thus takes the following form
\beq
E=E_d+E_g
\eeq
Fixation of a certain intra-Landau quantum number and all other eigenstates are of the type \cite{Ras,Sch}
\beq
|\psi_{d}\rangle={1\ov\sqrt{2}}|n_d\rangle+{1\ov\sqrt{2}}|n_d-1\rangle,\quad |\psi_{g}\rangle={1\ov\sqrt{2}}|n_g\rangle+{1\ov\sqrt{2}}|n_g-1\rangle.
\eeq
To find the system's spectrum, we begin with the Hamiltonian $H_g$, and the second may be obtained in the same manner, indeed
\beq
\left(
        \begin{array}{cc}
          \hbar\omega\left(a_g^\dag a_g+{1\ov2}\cosh \lambda\right)-E_g & \sqrt{2}\alpha{\hbar\ov l}e^{-i{\theta\ov2}}a_g^\dag \\
          \sqrt{2}\alpha{\hbar\ov l}e^{i{\theta\ov2}}a_g & \hbar\omega\left(a_g^\dag a_g+{1\ov2}\cosh \lambda\right)-E_g \\
        \end{array}
      \right)\left(
               \begin{array}{c}
                 |n_g\rangle \\
                |n_g-1\rangle \\
               \end{array}
             \right)
      =0
\eeq
then we obtain

\beq
\left\{
\begin{array}{r c c l}
\left(\hbar\omega\left(n_g+{1\ov2}\cosh \lambda\right)-E_g\right)+\sqrt{2}\alpha{\hbar\ov l}\sqrt{n_g}e^{-i{\theta\ov2}} &=& 0\\
\sqrt{2}\alpha{\hbar\ov l}\sqrt{n_g}e^{i{\theta\ov2}}+\left(\hbar\omega\left(n_g-1+{1\ov2}\cosh \lambda\right)-E_g\right) &=& 0\\
\end{array}
\right.
\eeq
by solving the equations system we get
\beq
E_g^\pm= \hbar\omega\left(n_g+\frac{1}{2}\left(\cosh\lambda-1\right)\pm\sqrt{{2\alpha^2\mu\ov\hbar\omega}n_g+1}\right)
\eeq

and by same way we can find the eigenvalue of $H_g$, so
\beq
E_d^\pm= \hbar\omega\left( n_d+\frac{1}{2}\left(\cosh\lambda-1\right)\pm\sqrt{{2\alpha^2\mu\ov\hbar\omega}n_d+1}\right)
\eeq
and the total eigenvalue of the system is
\beq
E=\hbar\omega\left(n_g+n_d+\cosh\lambda-1\pm\left(\sqrt{{2\alpha^2\mu\ov\hbar\omega}n_g+1}+\sqrt{{2\alpha^2\mu\ov\hbar\omega}n_d+1}\right)\right)
\eeq
\section{Connection to the Jaynes–Cummings model}
Recently, intriguing links between many branches of physics have been discovered. Between these is the special connection between condensed matter and high energy physics made possible by the combination of graphene and quantum electromagnetics, as cited by \cite{Kats}. This prompted us to seek more linkages and create additional connections between diverse systems. Thus, By showing how our system is linked to quantum optics via a mapping between the required Hamiltonian and the J–C model, we create a connection to another branch of physics. This may contribute to our understanding of many elements of quantum optics.
\subsection{Models equivalence}
To demonstrate the significance of our findings so far, using the J–C theory for an atomic shift in a radiation source with a frequency $\omega$, we show that our Hamiltonian is formally equivalent to two repetitions of the model. In order to do this, we use the same methodology as Ackerhalt and Rzazewski when studying operators' perturbation theory in the Heisenberg image \cite{Acke}. For the second situation, we deduce the appropriate results from our findings in a straightforward manner. To continue, we must reorganize our Hamiltonian such that each component is dealt with individually and the corresponding connection is established. To begin, we divide our Hamiltonian into two halves and simplify the mathematics by setting $\hbar=1$.

\beq
H=H_g+H_d
\eeq
in this case, the first half is
\beq\lb{H_g}
H_g=\omega a_g^\dag a_g+{\omega\ov2}\cosh\lambda+H_R^g
\eeq
and the second is as follows:
\beq\lb{hdd}
H_d=\omega a_d^\dag a_d+{\omega\ov2}\cosh\lambda+H_R^d
\eeq
To begin, let us examine the Hamiltonian (\ref{H_g}) and see how it relates to the Jaynes–Cummings model. In doing so, we demonstrate that the equation (\ref{H_g}) may be expressed as 
\beq\lb{H_gg}
H_g=\omega M_g+\alpha'\left(e^{-i{\theta\ov2}}a_g^\dag\sigma^++e^{i{\theta\ov2}}a_g\sigma^-\right)
\eeq
the two operators, $M_g $ and $\sigma^\pm$, are denoted by and respectively
\begin{eqnarray}
  M_g &=& a_g^\dag a_g+{1\ov2}\cosh\lambda \\
  \sigma^\pm &=& {1\ov2}\left(\sigma_x\pm i\sigma_y\right)
\end{eqnarray}
and we have assigned the constant $\alpha'$ to the value of
\beq
\alpha'=\left(\omega\mu\right)^{1\ov2}\alpha
\eeq
We present the Heisenberg equation of motion for the operators $a_g^\dag$ and $\sigma^-$ in order to investigate the dynamics associated with the Hamiltonian (\ref{H_gg}). They are as follows:
\begin{eqnarray}
  {d\ov dt}a_g^\dag &=& i[H_g,a_g^\dag] \\
  {d\ov dt}\sigma^- &=& i[H_g,\sigma^-] .
\end{eqnarray}
The result of a simple calculation is
\begin{eqnarray}
  \lb{vvvv} \left(i{d\ov dt}+\omega e^\lambda\right) a_g^\dag &=& -\alpha'e^{{1\ov2}\left(2\lambda+i\theta\right)}\sigma^- \\
  i{d\ov dt}\sigma^- &=& -\alpha'e^{-i{\theta\ov2}}\sigma_za_g^\dag  \lb{vvv}
\end{eqnarray}
using the formula
\beq
\sigma_z\sigma^-=-\sigma^-
\eeq
the equation (\ref{vvv}) is transformed into the following:
\beq\lb{ff}
i{d\ov dt}\sigma^- = \alpha'e^{-i{\theta\ov2}}a_g^\dag
\eeq
Above, we calculated the Heisenberg equations of motion for the operators $a^\dag$ and $\sigma^-$ using the Heisenberg equations of motion. As a result, we are looking for solutions in the form of
\beq\lb{sigma}
\sigma^-(t)=s_+^ge^{i\beta_+ t}+s_-^ge^{i\beta_- t},
\eeq
\beq\lb{aaa}
a_g^\dag(t)=d_+^ge^{i\beta_+ t}+d_-^ge^{i\beta_- t}
\eeq
where $r_{\pm}^g$, $d_{\pm}^g$ and $s_{\pm}^g$ are initial time operators. In order to get the auxiliary equation for $\beta_1$, we must first substitute a trial solution of the type $e^{i \beta t}$ into (\ref{ff}) and ((\ref{vvvv}) to get
\beq
\beta^2_1-\omega e^{\lambda}\beta_1+\alpha'^2e^{\lambda}=0
\eeq
it has two roots, which are represented as follows:
\beq
\beta^{\pm}_1=\Omega\pm \omega_g
\eeq
in this case, $\Omega={\omega e^{\lambda}\ov2}$, and the values $\omega_g$ are determined by the equation
\beq
\omega_g=  \sqrt{\Omega^2-2\mu\alpha^2\Omega}
\eeq
such as the operators constant  $s_{\pm}^g$ and $d_{\pm}^g$, which are established in terms of the initial time operators $a_0^\dag$ and $\sigma_0^-$. Indeed, by decomposing $a_g^\dag(t)$ and $\sigma^-(t)$, one can get the operator's constants $s_{\pm}$ and $d_{\pm}$ by setting $t = 0$ in (\ref{sigma}) and (\ref{aaa}). These establish the relationships
\begin{eqnarray}
  \sigma_0^- &=& s_{+}^g+s_{-}^g \\
  a_{g}^\dag(0) &=& d_{+}^g+d_{-}^g
\end{eqnarray}
If we replace the solutions (\ref{sigma}) and (\ref{vvvv}) with (\ref{vvvv}) and (\ref{ff}), respectively, and then evaluate these equations at $ t=0$, the initial operators $s_{\pm}^g$ and $d_{\pm}^g$ may be represented as
\beq
d_-^g={-\beta_1^-a_{g}^\dag(0)-\alpha'_\theta e^\lambda\sigma_0^-\ov 2r_g^+}
\eeq
\beq
d_+^g={\beta_1^+a_{g}^\dag(0)+\alpha'_\theta e^\lambda\sigma_0^-\ov 2r_g^+}
\eeq
\beq
s_-^g={\alpha'_{-\theta}  a_{g}^\dag(0)+\beta_1^+\sigma_0^-\ov 2r_g^+}
\eeq
\beq
s_+^g={-\alpha'_{-\theta}  a_{g}^\dag(0)+\beta_1^-\sigma_0^-\ov 2r_g^+}
\eeq
where $\alpha'_{\pm\theta}=\alpha' e^{\pm i{\theta\ov2}}$. After combining all of the variables, we get the final solutions of (\ref{vvvv}) and (\ref{ff}). They are as follows:
\beq
\sigma^-(t)={e^{i\Omega t}\ov \omega_g}\left(i\alpha'_{-\theta}\sin(\omega_gt)a_{g}^\dag(0)+\left(\Omega\cos(\omega_gt)-i\omega_g\sin(\omega_gt)\right)\sigma^-_0\right)
\eeq
and

\beq
a_g^\dag(t)={e^{i\Omega t}\ov \omega_g}\left(\left(i\Omega\sin(\omega_gt)+\omega_g\cos(\omega_gt)\right)a_{g}^\dag(0)+i\alpha'_{\theta}e^\lambda\sin(\omega_gt)\sigma^-_0\right)
\eeq
They are the precise operator solutions for the fundamental variables Jaynes–Cummings.

For the second part (\ref{hdd}), we are using the same method as previously to get comparable outcomes. In fact, the operator is introduced
\beq
M_d=a_d^\dag a_d+{1\ov2}\cosh\lambda
\eeq
we may write the Hamiltonian $H_d$ as follows:
\beq
H_d=\omega M_d-\alpha'\left(e^{-i{\theta\ov2}}a_d\sigma^++e^{i{\theta\ov2}}a_d^\dag\sigma^-\right).
\eeq
In the same manner, we come up with appropriate solutions
\beq
\sigma^+(t)={e^{i\Omega t}\ov \omega_d}\left(\alpha'_{\theta}\cos(\omega_dt)a_{d}^\dag(0)+\left(\Omega\cos(\omega_dt)+i\omega_d\sin(\omega_dt)\right)\sigma^+_0\right)
\eeq
and

\beq\lb{add}
a_d^\dag(t)={e^{i\Omega t}\ov \omega_d}\left(\left(-i\Omega\sin(\omega_dt)+\omega_d\cos(\omega_dt)\right)a_{d}^\dag(0)+i\alpha'_{-\theta}e^\lambda\sin(\omega_dt)\sigma^+_0\right)
\eeq
where the quantity $\omega_d$ is determined by the formula
\beq
\omega_d= \sqrt{\Omega^2+2\mu\alpha^2\Omega}
\eeq
Based on our insights, these procedures describe how the second copy of the J-C model should be produced. We conclude that our system is equivalent to two J–C type oscillating at two distinct frequencies. Finally, we should point out that we can simply get the dynamics of the positions from the solutions shown before. It is true that the formulations of the complex position $z$ in terms of the annihilation and creation operators are equivalent
\beq
z=e^{i{\theta\ov2}}\left(q_1+iq_2\right)={\ell\ov\sqrt{2}}\left(a_d^\dag+a_g\right)e^{i{\theta\ov2}}
\eeq
If we assume that the starting velocity is zero at time $t=0$, we may simplify the formula for $z$. In this case, we get the following equation:
\beq
a_d^\dag(0)=a_g(0)
\eeq
then the complex position $z$ is
\begin{eqnarray}
 z(t) &=& {\ell\ov\sqrt{2}}\bigg({e^{i\Omega t}\ov \omega_d}\left(\left(-i\Omega\sin(\omega_dt)+\omega_d\cos(\omega_dt)\right)a_{d}^\dag(0)+i\alpha'_{-\theta}e^\lambda\sin(\omega_dt)\sigma^+_0\right) \\ \nonumber
   &+& {e^{-i\Omega t}\ov \omega_g}
   \left(\left(-i\Omega\sin(\omega_gt)+\omega_g\cos(\omega_gt)\right)a_{g}(0)-i\alpha'_{-\theta}e^\lambda\sin(\omega_gt)\sigma^+_0\right)
  \bigg)e^{i{\theta\ov2}}
\end{eqnarray}
The next section will be devoted to studying the correlation function for the absorption or emission of photons in order to provide context for our prior investigation.

\section{Correlation functions}
The benefit of having calculated operators' time evolution is that it is now simple to calculate significant amounts, such as $\zeta$, which in the Schrodinger image are difficult to calculate. For instance, kindly let us construct using (\ref{add}) and its Hermitian conjugate, the correlation function to destroy a photon at $t'$ time and create a photon at $t$ time:
\beq
\zeta=\langle a_d^\dag\left(t\right) a_d \left(t'\right)\rangle
\eeq
then we have
\begin{eqnarray}\lb{zzz}
  \zeta &=& \langle {1\ov {\omega_d}^2}\left(\left(-i\Omega\sin(\omega_dt)+\omega_d\cos(\omega_dt)\right)a_{d}^\dag(0)+i\alpha'_{-\theta}e^\lambda\sin(\omega_dt)\sigma^+_0\right) \\\nonumber
   &\times& \left(\left(i\Omega\sin(\omega_dt')+\omega_d\cos(\omega_dt')\right)a_{d}(0)-i\alpha'_{\theta}e^\lambda\sin(\omega_dt')\sigma^-_0\right)\rangle
\end{eqnarray}
In the Heisenberg picture, we can calculate the
quantity (\ref{zzz}) by looking at the time evolution of
the expectation value of the photon number operator, where absorption (emission) of a photon corresponds to the choice of initial state $|\psi(0)\rangle = |n_d, +\rangle (|\psi(0)\rangle= |n_d -1, -\rangle)$. Then  the expression of correlation function for absorption of photons by setting $t'=t$ in (\ref{zzz}) is written by
\beq\lb{za}
\zeta_A=\left(\left({\Omega\ov \omega_d}\right)^2\sin^2(\omega_dt)+\cos^2(\omega_dt)\right)n-\left({\alpha'e^{\lambda}\ov \omega_d}\right)^2\sin^2(\omega_dt)
\eeq
in addition to photon emission in the same way
\beq\lb{ze}
\zeta_E=\left(\left({\Omega\ov \omega_d}\right)^2\sin^2(\omega_dt)+\cos^2(\omega_dt)\right)\left(n-1\right)+\left({\alpha'e^{\lambda}\ov \omega_d}\right)^2\sin^2(\omega_dt)
\eeq
In (\ref{za}), the atom is initially in the ground state
such that only absorption of a photon is possible.
The expectation value of $a\dag(t)a(t)$ is therefore always less than or equal to the number of photons
at time $t = 0$. In (\ref{ze}), the atom is initially excited such that only emission of a photon is possible. The expectation value of $a\dag(t)a(t)$ is therefore always greater than or equal to the number of photons at time $t =0$. The probability of absorption
or emission can be identified as the magnitude of the second term on the right-hand side of equations. The probability of emission or absorption of a photon $P(t)$  \cite{Acke} is given by
\beq
P(t)=\left({\alpha'e^{\lambda}\ov \omega_d}\right)^2\sin^2(\omega_dt)
\eeq
\subsection{Strong and weak coupling}
\subsubsection{weak Rashba coupling}
in the case when the coupling is very small $\alpha\rightarrow0$ that is to say that $\omega_d=\Omega$, then the expression of the correlation function for emission case is written by
\beq
\zeta=n-{4\mu\alpha^2\ov \omega}\sin^2\left({\omega e^\lambda\ov 2}t\right).
\eeq
It is clear from the previous equation that $\zeta$ is affected by many variables. The study of this function may help us understand the correlation behavior of our system. The correlation function for the emission in this instance is now shown to allow us to draw some inferences.  Assume that $\alpha=0.1,0.2,0.4$ are fixed values. The below curves represents the variation of $\zeta$  in terms of $\lambda$ parameter.
\begin{figure}[H]
  \centering
  \includegraphics[width=7 in]{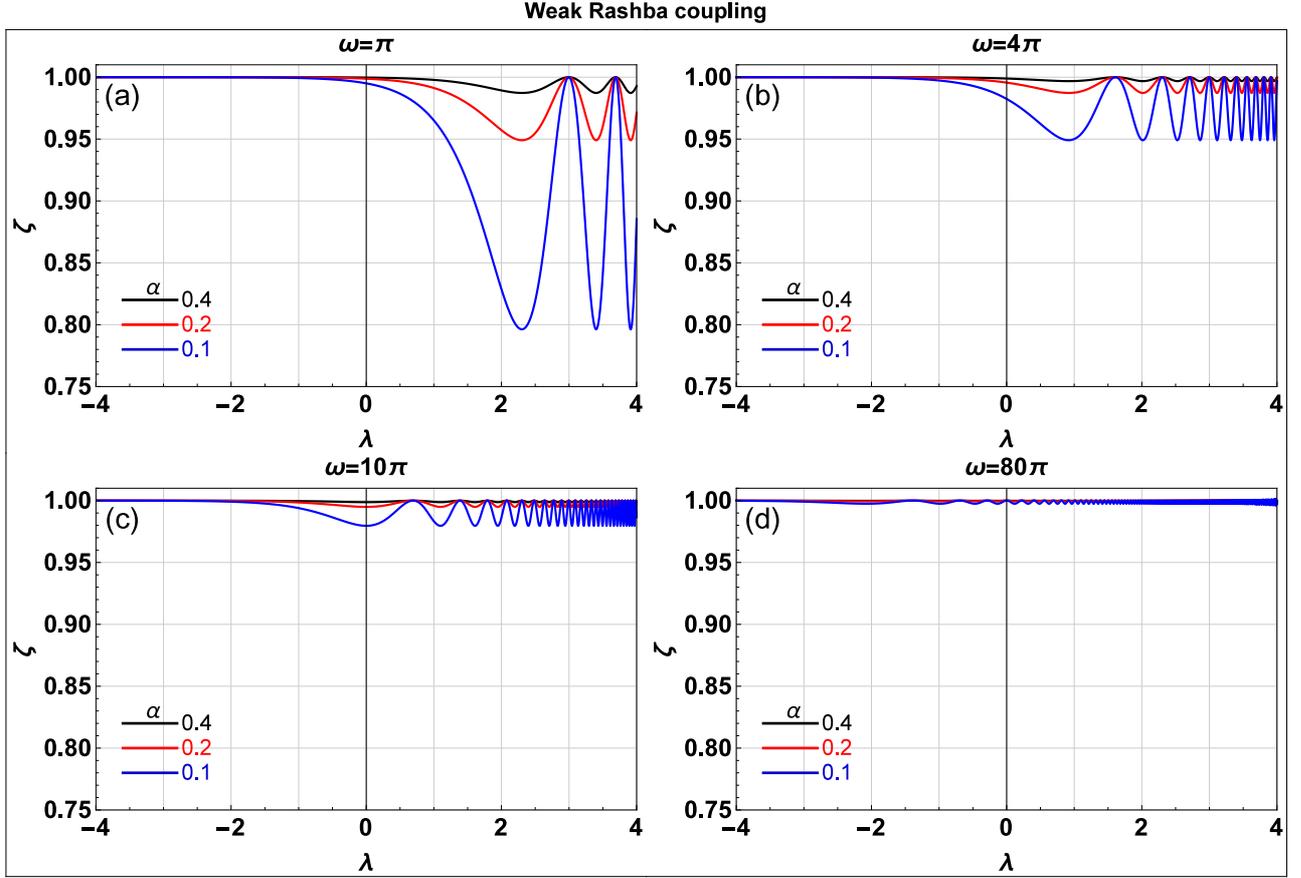}\\
  \caption{The correlation function  in terms of $\lambda$ parameter for the first Landau level at a given instant for weak coupling for the values $\omega=\pi,4\pi,10\pi,80\pi$ and $\alpha=0.1,0.2,0.4$. }\label{co}
\end{figure}

We remark  in figure (\ref{co}) in the first hand that the correlation function for weak Rashba coupling remains almost
constant $\zeta\sim1$ in high frequency regime $(\omega=80\pi)$ for any value of coupling, and the other hand for low frequency regime, the amplitude
of the oscillation of the correlation function decreases for the large values of the coupling. The seconde remak is that $\zeta$ is always positive, which shows that the destruction and creation of photons in this case increase or decrease together.

\subsubsection{Strong Rashba coupling}
in the case when Rashba coupling is strong the frequency $\omega_d$ is written by
\beq
\omega_d\simeq\alpha\sqrt{\mu\omega e^\lambda}
\eeq
with this approximation the correlation function become
\beq
\zeta=\left({\omega e^\lambda\ov4\alpha^2\mu}\sin^2\left(\alpha\sqrt{\mu\omega e^\lambda} t\right)+\cos^2\left(\alpha\sqrt{\mu\omega e^\lambda}t\right)\right)n-e^\lambda\sin^2\left(\alpha\sqrt{\mu\omega e^\lambda} t\right)
\eeq
The graphs below illustrate the variation of $\zeta$ in terms of the $\lambda$ parameter for Rashba coupling values $\alpha=5,10,15$.
\begin{figure}[H]
  \centering
  \includegraphics[width=7 in]{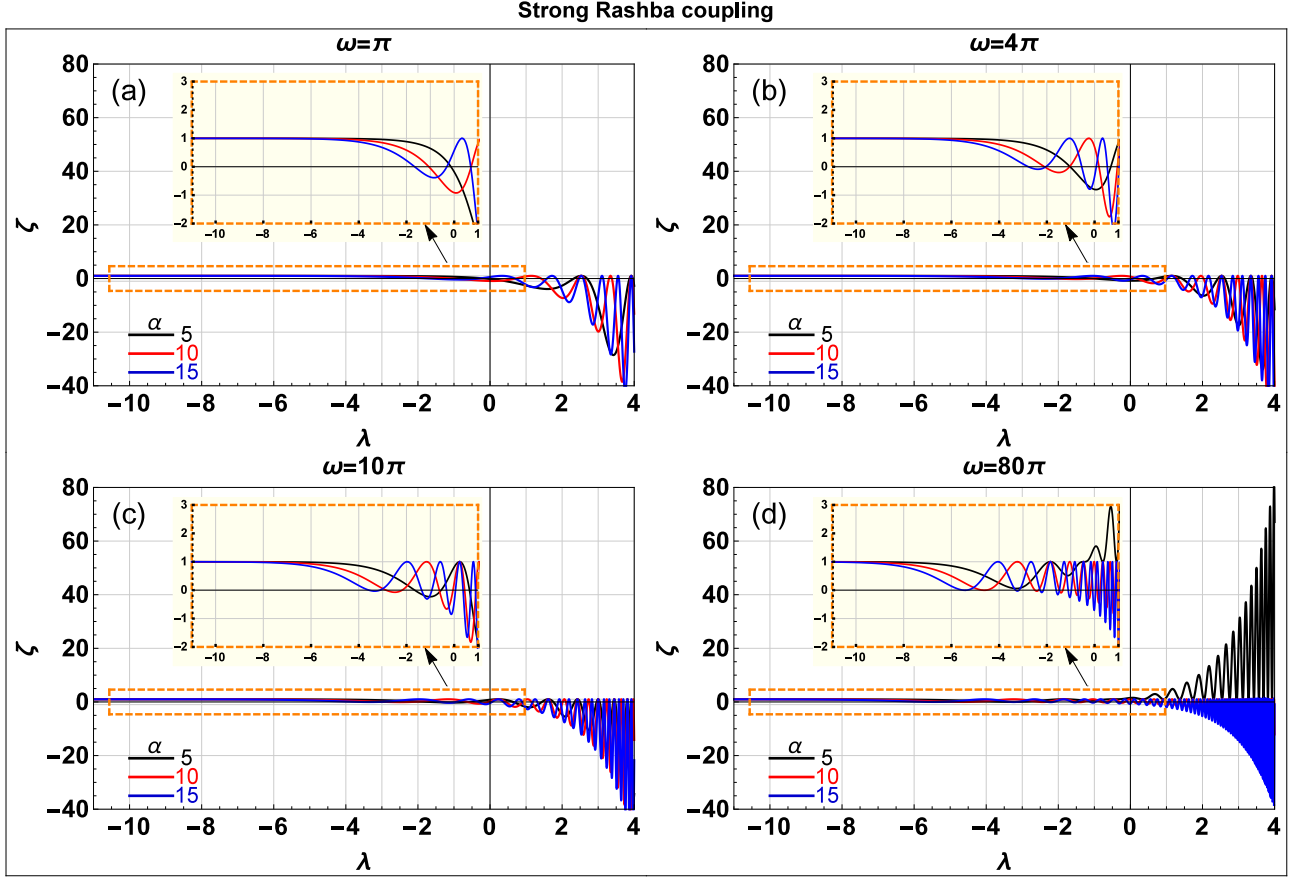}\\
  \caption{The correlation function  in terms of $\lambda$ parameter for the first Landau level at a given instant for strong coupling for the values $\omega=\pi,4\pi,10\pi,80\pi$ and $\alpha=5,10,15$.}\label{corr}
\end{figure}
The figure (\ref{corr}) shows that the correlation function of photon creation and annihilation oscillates with a variable frequency depending on $ \lambda $ and that the oscillation amplitude also increases with large values of the parameter $ \lambda $ that of on the one hand and on the other hand it depends on the coupling and the totality of $ \zeta $ is negative, which shows that there is an antagonistic variation between the destruction and the creation of the photons.

\section{Conclusion}
We have looked into the fundamental properties of two anharmonic oscillators with Rashba spin–orbit coupling and discovered some interesting results.
In this work, we demonstrated that it is feasible to get a diagonalized Hamiltonian by scaling various variables.
In fact, this is accomplished via the application of a unitary transformation.
Essentially, this was beneficial in that it allowed us to get the eigenvalues and their associated wavefunctions in a straightforward manner by considering the coupling parameter $\alpha$.
This method has specifically been used to the splitting of the related Hamiltonian into halves. This decomposition was advantageous in that it allowed for the generation of various spectra and the discovery of complete solutions for the energy spectrum.
In order to establish a link with quantum optics, we have developed a technique that is based on deriving two Hamiltonians from a single Hamiltonian.
Furthermore, we have also shown that by dividing the Hamiltonian into halves, it is feasible to reconstruct the J–C model, which represents a two-state system.
This was accomplished by computing the dynamics of the rising Pauli operator $\sigma^+$  and the creation operators $a^\dag_d,a^\dag_g$ using the Heisenberg dynamics.
We arrived at the appropriate operator solutions for the fundamental J–C variables by evaluating a multitude of characteristic equations.
After that, we looked at the behaviors of the correlation function of emission and absorption photons under strong and weak coupling to see how it influences our correlation function.


\section*{Acknowledgment}


We greatly appreciate Mohamed Monkad, director of the Laboratory of Physics of Condensed Matter (LPMC), Faculty of Sciences, Choua\"ib Doukkali University, for his generous donations. And we would like to express our gratitude to El Mouhafid Abderrahim for his numerical assistance.


\end{document}